\documentclass[epj,twocolumn]{webofc}
\usepackage[varg]{txfonts}   

\usepackage{graphicx}
\usepackage{amsmath}   
\usepackage{float}
\usepackage{subfig}

\wocname{MPGD,2015}

\woctitle{4th Conference on Micro-Pattern Gaseous Detectors}

\begin{document}


\title{A novel application of Fiber Bragg Grating (FBG) sensors in MPGD}

\author { D.~Abbaneo \inst{19}
  \and M.~Abbas \inst{19}
  \and M.~Abbrescia \inst{2}
  \and M.~Abi Akl \inst{14}
  \and O.~Aboamer \inst{8}
  \and D.~Acosta \inst{17}
  \and A.~Ahmad \inst{21}
  \and W.~Ahmed \inst{21}
  \and A.~Aleksandrov \inst{31}
  \and P.~Altieri \inst{2}
% ...
}

\institute{  RWTH Aachen University, III Physikalisches Institut A, Aachen, Germany
  \and
  INFN Bari and University of Bari, Bari, Italy
  \and
  Peking University, Beijing, China
  \and
  INFN Bologna and University of Bologna, Bologna, Italy
  \and 
  Universite Libre de Bruxelles, Brussels, Belgium
  \and
  National Institute of Science Education and Research, Bhubaneswar
  \and
  Institute for Particle and Nuclear Physics, Wigner Research Centre for Physics, Hungarian Academy of Sciences, Budapest, Hungary
  \and
  Academy of Scientific Research and Technology - Egyptian Network of High Energy Physics, ASRT-ENHEP, Cairo, Egypt
  \and
  Helwan University \& CTP, Cairo, Egypt
  \and
  Texas A\&M University, College Station, U.S.A.
  \and
  Institute for Nuclear Research of the Hungarian Academy of Sciences (ATOMKI), Debrecen, Hungary
  \and
  University of Delhi, Delhi, India
  \and
  Wayne State University, Detroit, U.S.A
  \and
  Texas A\&M University at Qatar, Doha, Qatar
  \and
  ENEA - Frascati Research Centre, Frascati RM- Italy
  \and
  Laboratori Nazionali di Frascati - INFN, Frascati, Italy
  \and
  University of Florida, Gainesville, U.S.A.
  \and
  Petersburg Nuclear Physics Institute, Gatchina, Russia
  \and
  CERN, Geneva, Switzerland
  \and
  Ghent University, Dept. of Physics and Astronomy, Ghent, Belgium
  \and 
  National Center for Physics, Quaid-i-Azam University Campus, Islamabad, Pakistan
  \and
  Saha Institute of Nuclear Physics, Kolkata, India
  \and
  Lappeenranta University of Technology, Lappeenranta, Finland
  \and
  University of California, Los Angeles, U.S.A.
  \and
  Florida Institute of Technology, Melbourne, U.S.A.
  \and
  Bhabha Atomic Research Centre, Mumbai, India
  \and
  INFN Napoli, Napoli, Italy
  \and
  INFN Pavia and University of Pavia, Pavia, Italy
  \and
  University of Rome “La Sapienza” (IT) - Facoltà di Ingegneria,
  Ingegneria Chimica Materiali ed Ambiente, Roma, Italy
  \and
  IRFU CEA-Saclay, Saclay, France
  \and
  Institute for Nuclear Research and Nuclear Energy, Sofia, Bulgaria
% ...
}


\label{sec-2}
\abstract{
We present a novel application of Fiber Bragg Grating (FBG) sensors in
the construction and characterisation of Micro Pattern Gaseous
Detector (MPGD), with particular
attention to the realisation of the largest triple (Gas electron
Multiplier) GEM chambers so far operated, the GE1/1 chambers of the CMS experiment at LHC. The GE1/1
CMS project consists of 144 GEM chambers of about 0.5 m$^2$ active area
each, employing three GEM foils per chamber, to be installed in the
forward region of the CMS endcap during the long shutdown of LHC in
2108-2019. The large active area of each GE1/1 chamber consists of GEM
foils that are mechanically stretched in order to secure their
flatness and the consequent uniform performance of the GE1/1 chamber
across its whole active surface. So far FBGs have been used in high
energy physics mainly as high precision positioning and re-positioning
sensors and as low cost, easy to mount, low space consuming
temperature sensors. FBGs are also commonly used for very precise
strain measurements in material studies. In this work we present a
novel use of FBGs as flatness and mechanical tensioning sensors
applied to the wide GEM foils of the GE1/1 chambers. A network of FBG
sensors have been used to determine the optimal mechanical tension
applied and to characterise the mechanical tension that should be
applied to the foils. We discuss the results of the test done on a
full-sized GE1/1 final prototype, the studies done to fully
characterise the GEM material, how this information was used to define
a standard assembly procedure and possible future developments.

 }

\maketitle

\section{Introduction}
\label{intro}

\begin{figure}
\centering
\includegraphics[width=0.9\linewidth]{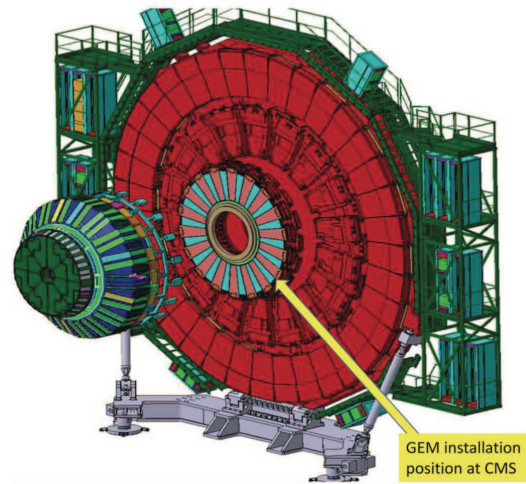}
\caption{GEM detector installation position in CMS \cite{Colaleo:2021453}}
\label{fig:CMS}
\end{figure}

\begin{figure}[ht]
  \centering
    \includegraphics[width=0.4\textwidth]{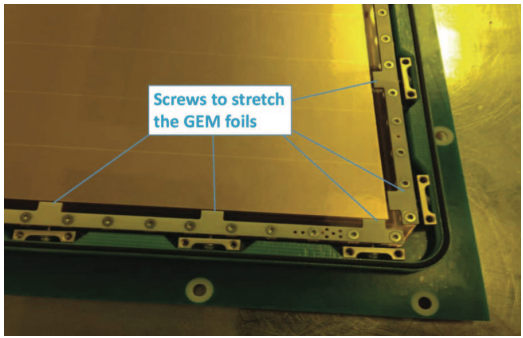}
  \caption{Lateral screws for stretching the GEM foils stack.}
    \label{fig:subfig1}
\end{figure}

To upgrade the Compact Muon Solenoid (CMS\cite{Chatrchyan:2008aa}) 
muon system 144 GEM chambers will be installed in the high
pseudo-rapidity region of CMS during Long Shutdown 2 (LS2) of the Large Hadron Collider
\cite{Abbaneo:2014zxc}. The GEMs can provide extra leverage on
precision studies of standard model physics, as well as open up a
window to explore exotic signatures with muons in the high
pseudorapidity region
\cite{Abbaneo:2014lja}.  The GEM chambers will installed, as shown in fig. \ref{fig:CMS},
very close to the beam pipe where a high flux of low Pt muons is
expected.  The GEM chambers can easily handle this rate due to their
high rate capability of 100 MHz/cm$^2$.  The large active area of each
GE1/1 (GEM Endcap) chamber, approximately 0.4 m$^2$
\cite{Colaleo:2021453}, consists of a triple-GEM foil stack stretched
by means of screws placed around the stack frame (fig.\ref{fig:subfig1}). These
foils need to be stretched simultaneously in order to secure the
planarity and consequent uniform performance of the GE1/1 chamber
\cite{Abbaneo:2013nba}. The GE1/1 detector technology used for CMS is
described in detail somewherelse \cite{Gilles}. 
\\* The FBG sensors
act as low cost precision spatial and temperature sensing tools and
they are commonly used for strain measurements \cite{Benussi:2010yw}
\cite{Caponero:2012py} \cite{Benussi2012483}. These sensors are
commercially available and have relative low cost. In this work
FBG sensors are used to measure the planarity and mechanical tension
of the GEM foils in the GE1/1 chambers. The GE1/1 assembly procedure
employs a mechanical stretching procedure to apply tension to the GEM
foils by means of a series of lateral screws inserted into the
internal GE1/1 frame. This technology allows mechanical assembly of
the GEM chamber without the use of internal spacers or glue. 

\section{FBG sensors as a strain measurement}
A FBG is a type of distributed Bragg reflector, constructed in a short
segment of optical fiber that reflects particular wavelengths of light 
and transmits all others. The sensitivity of FBG in terms of strain, 
defined as relative elongation w.r.t. the initial position is of the
order of 0.1 $\mu$. This is achieved by creating a periodic variation
 in the refractive index of the fiber core, which generates a
 wavelength-specific dielectric mirror. Therefore it can be used as a 
strain measurement tool since variation of the FBG translates into 
different light frequency response.

\begin{figure}[ht]
  \centering
    \includegraphics[width=0.3\textwidth]{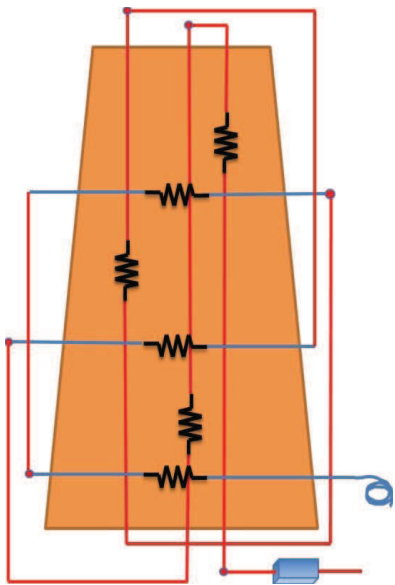}
    \caption{Conceptual schema of the FBG sensors connection on one GEM foil.}
    \label{fig:subfig2}
\end{figure}

In order to validate the mechanical stretching technique a network of
FBG sensors is affixed on the triple-GEM stack as shown in figure
\ref{fig:subfig2}. Each sensor is glued on the GEM foil using a very thin layer of
epoxy glue. Due to the high temperature sensitivity of the FBG the
thermal dilatation of the materials used are corrected by using a
separate FBG sensor on a 10$\times$10 cm$^2$ GEM detector.  The test
is performed by modifying the stretching conditions of the GEM foils
stack with real time monitoring and recording of the FBG sensors data.
The test starts with the chamber normally assembled with the GEM stack
mechanically stretched to the nominal tensile load.  After some time
while steady in the starting condition, the mechanical stretching of
the GEMs is released and kept in such condition for several hours.
Finally the GEMs are stretched again up to the nominal tensile
load. The trends of the FBG sensors are shown in figure \ref{fig:subfig3}.

The steep variations of the strain evident in figure \ref{fig:subfig3} correspond to
the actions of un-screwing and screwing the mechanical stretchers
during the test. The initial stretch value is assumed as reference
condition with strain = 0. When stretchers are un-screwed the strain
goes to the lower value, different strain values apply to different
foils as they fold quasi-free and assume unequal conditions. After the
stretchers are screwed back, the strain value is similar for all
foils, showing that they all experience similar stretching, about the
original value of the reference condition. Thus it can be inferred
that at the predetermined tensile load all foils reach a similar
stretched level although they started from different values. From the
plot it can be seen that all the sensors of the network react at the
same moment. These results allow us to validate the mechanical
stretching assembly technique for GE1/1 chambers. Further tests are
ongoing to confirm other important parameters such as the optimal
tensile load to be applied to the GEMs and the maximum planarity
obtainable for the GEMs without applying a load beyond the Young's
region for GEM foils.

\section{Another FBG application: the load gauge}
\begin{figure*}[ht]
  \centering
  \subfloat[Subfigure 1 list of figures text][ ]{
    \includegraphics[width=0.45\textwidth]{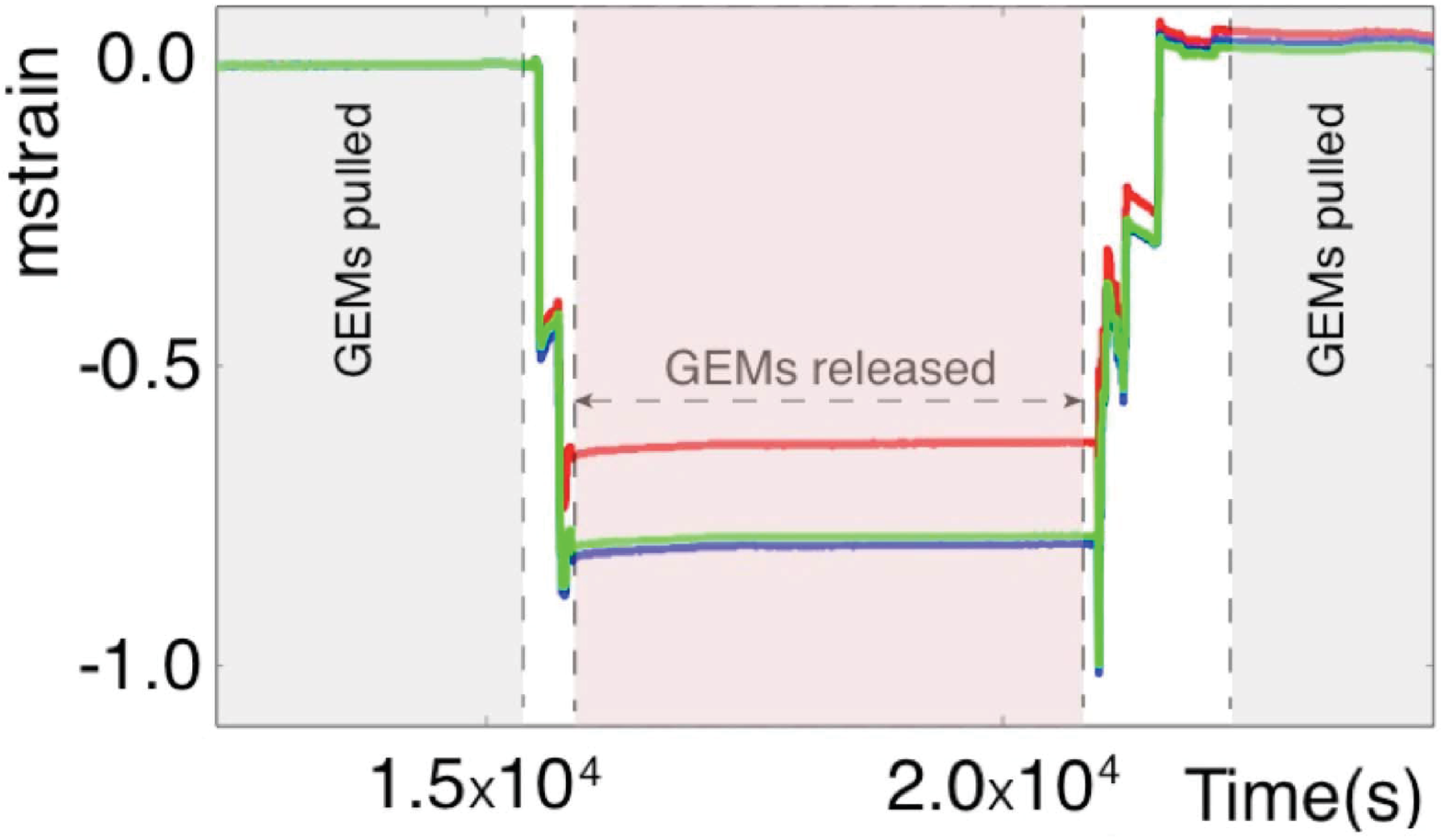}
    \label{fig:subfig3}}
  \qquad
  \subfloat[Subfigure 2 list of figures text][ ]{
    \includegraphics[width=0.45\textwidth]{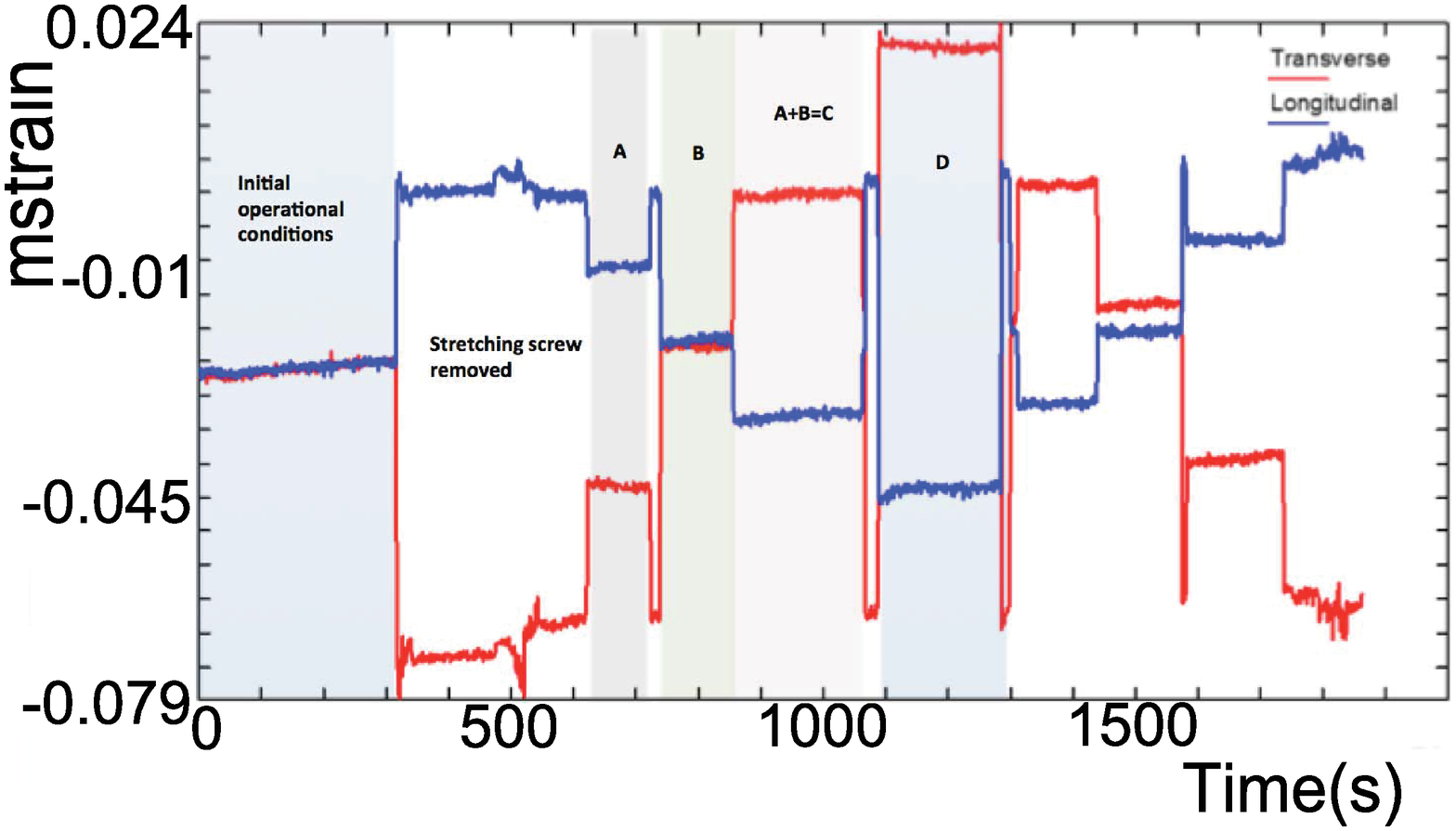}
    \label{fig:subfig4}}
  \caption{(a) Three regions corresponding to the mechanical
    stretched, loose and again stretched triple GEM foils stack
    respectively. (b) The response of FBG sensors during the test as load gauge}
\end{figure*}

Another important application of FBG in GEM chamber 
construction is the possibility to be used as load gauge for precise measurement 
of the tensile load applied to the foils, of the different layers, in the same moment.
This is extremely important in the case of the GE1/1 chambers since, 
during their assembly procedure, the foils are stretched by means of
screwing nuts with dynamometric screwdriver. It is thus very important
to know precisely how much to pull the foils in order to avoid to 
stretch them too much, reaching the mechanical load with will cause the GEM foils to be
operated outside their elastic range (Young region).
In order to demonstrate the idea of FBG as load gauge we have done the
following test which results in terms of FBG response are shown in
figure \ref{fig:subfig4}. After having properly stretched the GE1/1 GEM stack, we
have removed a single stretching screw from the chamber and
replaced it with an eyelet screw on which we have fixed 
a stainless steel wire used to attach different weights. The idea was
to add weights in different steps till they applied load reproduce in
the FBG facing the eyelet the same response when the GEMs are stretched
by means of the original screw. 

In this test few sample of weights (lead bricks) are used. 
We took 4 points corresponding to the following 
weights A=2.86 kg, B=5.65 kg, C=8.5 kg, D=11.3 kg. Fig
\ref{fig:subfig4} shows the response of the FBG during
this test. The RED and BLUE lines correspond to the two sensors
closest to the point where the load is applied (RED sensor parallel to the
line of the stainless steel wire, BLUE orthogonal to the wire
direction). In the plot are visible different regions corresponding to
the different actions performed to the foils. The first flat region
correspond to the initial position of the foils stack,
meaning the GEMs properly stretched with the original screw. Then the screw is
removed and the two sensors immediately record the effect on the foils
tension. There are some movements in the following flat regions
due to the handling during the substitution of the screw. The first
weight A applied shows its effect on the sensors. Then weight A is 
removed to load weight B (it is visible the short foils relaxation
during this action). The third step was to add the weight A to B and
 after removing A+B (weight C) we add weight E. Finally we redo
 the same steps backwards as shown in the plot, removing the weights
in steps. 
By analysing the plots shown in fig. \ref{fig:subfig4} we were able to
determine the load applied on a single screw once the GEM stack is
stretched to its operational value. This value is of the order 5.65 kg 
and is very close to the expected value of 3 kg/cm considering that
the crimping length of one screw is 2 cm. This result is important
since other measurements presented in this conference shown that the
GEM foils Young region ends around 8 kg/cm. This value ensures
that the mechanical tension applied to GEM foils is still in the
elastic regime by a wide margin.

\section{Conclusion}
With the use of FBG sensors we successfully demonstrated that the
novel glue-less technique adopted to assemble the GE1/1 chambers for
the LS2 update of CMS is reliable and guarantees the correct tensioning
of the three GEM foils. By applying the correct tension across the 
GEM stack the uniform gaps spacing is obtained, which is extremely
important to get the required performance of the detector. 
We have also demonstrated the FBG are excellent load gauges 
that can be used to determine very precisely the applied stretching
force to GEM foils which is a critical point for the correct assembly
of the GE1/1 chambers and to secure their homogeneity.
Other tests are ongoing by using the same FBG sensors to optimise the
tensile load in order to avoid damage and guarantee planarity of the
GEM foils.

\section* {Acknowledgments}
We gratefully acknowledge the support of FRS-FNRS
(Belgium), FWO-Flanders (Belgium), BSF-MES (Bulgaria),
BMBF (Germany), DAE (India), DST (India), INFN (Italy),
NRF (Korea), QNRF (Qatar), and DOE (USA). 
This project has received funding from the European Union’s Horizon
2020 Research and Innovation programme under Grant Agreement
no. 654168.

\end{document}